\documentclass[final,3p,times]{elsarticle}

\usepackage{graphicx}
\usepackage{amssymb}

\usepackage{lineno}
\usepackage{graphicx}
\usepackage{float}
\usepackage{longtable,multicol,relsize }
\usepackage{epsfig,amsfonts,amsmath}
\usepackage{bm}

\usepackage{caption}
\usepackage{subcaption}
\usepackage{amsmath}
\usepackage{wasysym}
\usepackage{bigints}

\usepackage{lineno}
\usepackage{epsfig,amsfonts,amsmath}
\usepackage{wrapfig}
\usepackage{breqn,soul}
\usepackage[colorinlistoftodos,prependcaption,textsize=tiny]{todonotes}

\def\correspondingauthor{\footnote{Corresponding author.}}

\newcommand{\vect}[1]{\boldsymbol{\mathit{#1}}}

\renewcommand{\det}{\mathrm{det}}

\newcommand{\tens}[1]{\mathbf{#1}}

\DeclareMathOperator*{\argmax}{arg\,max}

\journal{Journal Name}

\begin{document}

\begin{frontmatter}


\title{A Bayesian Surrogate Constitutive Model to Estimate Failure Probability of Rubber-Like Materials}

\author[label1,label2]{Aref Ghaderi}
\author[label1,label3]{Vahid Morovati}

\author[label1,label4]{Roozbeh Dargazany\correspondingauthor{}}

\address[label1]{Department of Civil and Environmental Engineering, Michigan State University}
\address[label2]{ghaderi1@msu.edu}
\address[label3]{morovati@msu.edu}
\address[label4]{roozbeh@msu.edu}

\begin{abstract}
{In this study, a stochastic constitutive modeling approach for elastomeric materials is developed to consider uncertainty in material behavior and its prediction. This effort leads to a demonstration of the deterministic approaches error compared to probabilistic approaches in order to calculate the probability of failure. First, the Bayesian linear regression model calibration approach is employed for the Carroll model representing a hyperelastic constitutive model. The developed model is calibrated based on the Maximum Likelihood Estimation (MLE) and Maximum a Priori (MAP) estimation. Next, a Gaussian process (GP) as a non-parametric approach is utilized to estimate the probabilistic behavior of elastomeric materials. In this approach, hyper-parameters of the radial basis kernel in GP are calculated using L-BFGS method. 
To demonstrate model calibration and uncertainty propagation, these approaches are conducted on two experimental data sets for silicon-based and polyurethane-based adhesives, with four samples from each material. These uncertainties stem from model, measurement, to name but a few. 
Finally, failure probability calculation analysis is conducted with First Order Reliability Method (FORM) analysis and Crude Monte Carlo (CMC) simulation for these data sets by creating a limit state function based on the stochastic constitutive model at failure stretch. Furthermore, sensitivity analysis is used to show the importance of each parameter of the probability of failure. Results show the performance of the proposed approach not only for uncertainty quantification and model calibration but also for failure probability calculation of hyperelastic materials.}

\end{abstract}


\end{frontmatter}

\section{Introduction}

\paragraph{}
Nowadays, the application of rubber-like materials in several industries such as automotive, shipbuilding, and structural science, to name but a few, leads to a considerable investigation on the modeling of their properties. One of the most challenging problems for modeling the mechanical behavior of rubber-like materials is computational efforts. Elastomers are wide meshed cross-linked polymers that behave entropically elastic and do not show reversible deformation. They are usually classified as filled and unfilled categories, where in most cases, fillers are used to reinforce polymers.

\paragraph{}
The behavior of elastomers and rubber-like materials shows a non-linear trajectory, especially during large deformation. Hyperelastic constitutive models describe their behavior in small and large deformation. In these models, stress-strain relation is derived based on the strain energy function. Hence, researchers spare no effort to find a strain energy function that captures the behavior of elastomers under different loading states. The development of constitutive models for rubber-like materials is hindered by both incompleteness of the theoretical approach and limitation in experiment observation.

\paragraph{}
In the past decades, researchers have proposed many deterministic constitutive models based on materials' average response to mechanical elangation, in which most of them do not consider uncertainty. Thus, they cannot consider confidence bounds to estimate the response of the model or reduce the error of the model. In addition, one of the main challenges in deterministic approaches is the variation of failure stretch/stress of different samples, which cannot be defined in deterministic methods. So, failure should be defined as a Fuzzy variable rather than a binary process.

Uncertainty quantification (UQ) adds bounds on the behavioral prediction of the model. UQ plays a pivotal role in modeling under the framework of continuum mechanics. UQ can determine a mathematical model to calculate the error bounds. In computational modeling, due to the nature of data used, predictions are usually deterministic. In these methods, a single estimate is calculated based on the average response of the system. Although, in real problems, model prediction for a specific group of parameters and different combinations of parameter values of the model may have similar results \cite{honarmandi2019materials, vahidi2020memory}. One of these combinations is determined by a deterministic approach. However, probabilistic modeling can calculate different combinations of parameters in the form of the probability distribution of model prediction through the propagation of uncertainty. Another importance of UQ is the prediction of the failure of the material, which has a crucial role in the context of the safety factor in design. Also, a better estimation of safety factors leads to a cost reduction, resulting in generating confidence bounds in probabilistic modeling. So, failure can be seen as a probability, not a black or white occurrence. The sources of uncertainty, generally, are categorized based on their capability in uncertainty reduction: (i) "epistemic," the error reducible due to lack of knowledge, which can be reduced by cost, and (ii) "aleatory," the inherent error of the system, which we cannot or do not know how to reduce. The goal of UQ in computational modeling is the calculation of uncertainty in the modeling and its prediction. Two statistical views usually evaluate the quantification process: the frequentist view, which defines probability during a long-term observation based on the rate of occurrence, and the Bayesian view, which considers the degree of belief based on the combination of prior knowledge and new data for probability. So, in the frequentist view, parameters are fixed random variables. However, in the Bayesian perspective, parameters are random variables based on the available data. Although incorrect prior knowledge can mislead the model, the true definition is helpful in statistical inference.

\subsection{Past Research}

In 1948, Rivlin \cite{rivlin1948large} investigated fundamental concepts of large elastic deformation of isotropic materials. Hart-Smith \cite{hart1966elasticity} worked on finite deformation of rubber-like materials to calculate elasticity parameters for them. Ogden \cite{ogden1972large} proposed a model for large deformation to remove isotropic and incompressibility assumptions from the constitutive model. James et al. \cite{james1975strain}, in their work, proposed two analytic forms of the strain energy function for isotropic and incompressible materials. In 1997, Yeoh and Fleming \cite{yeoh1997new} tried to combine concepts of the statistical and phenomenological approach to suggest a constitutive model for rubber vulcanizates. Lambert-Diani and Rey \cite{lambert1998elaboration}, in 1998, proposed a family of functions of strain energy associated with the hyperelastic behavior of elastomers. In 1999, Yeoh \cite{yeoh1990characterization} proposed a new cubic strain energy function based on the first deformation invariant. After that, Pucci and Saccomandi \cite{pucci2002note} showed that the Gent model is similar to the 8-chain model with limited chain extensibility. They proposed a model that, with a minimum number of coefficients, can capture Treloar data \cite{treloar1975physics}. After one year, Beda and Chevalier \cite{beda2003hybrid} combined the Gent model and Ogden model to capture experimental data and mechanical behavior of rubber-like materials. They enhanced their models in the next years \cite{beda2005reconciling, beda2007modeling}. In 2011, Kroon \cite{kroon2011} proposed a model based on the eight-chain model by adding some topological constraint of moving space of chain. They showed the performance of their model with the experimental data set. Farhangi et al. investigated effect of fiber reinforced polymer tubes filled with recycled materials \cite{farhangi12020effect, farhangi2020effect}. Izadi et al. investigated effect of nanoparticles on mechanical properties of polymers \cite{izadi2019plasma, sinha2019novel, izadi2020mechanical}. Nunes and Moreira \cite{nunes2013simple}, in 2013, published a work that analyzes the simple shear state of an incompressible hyperelastic solid under large deformation by experimental and theoretical approaches. In 2016, Nkenfack et al. \cite{nkenfack2016hia, nguessong2016hia} proposed a new model by adding an integral density and an interleaving constraint part to the eight-chain model. In 2020, we proposed a phsics-based data-driven constitutive model for cross-linked polymers by embedding neural networks in micro-sphere \cite{ghaderi2020physics}. All of the introduced models above have a deterministic approach.

In recent years, several studies have been conducted on the stochastic modeling of constitutive models for rubber-like materials. In 2015, a Bayesian approach was employed for calibration of the constitutive model for soft tissue \cite{madireddy2015bayesian}. Recently, Brewick and Teferra \cite{brewick2018uncertainty} investigated on uncertainty quantification of the constitutive model for brain tissue. They consider the Ogden model as the reference model in their work. Kaminski and Lauke \cite{kaminski2018probabilistic}, in 2018, worked on probabilistic aspects of rubber hyperelasticity. They considered some basic models, from Neo-Hookean to Arruda-Boyce, and showed probabilistic characteristics, such as expectation, variance, skewness, and kurtosis. Also, recently, Mihai et al. \cite{fitt2019uncertainty} published their study on the uncertainty quantification of elastic materials. Another study has been used as a Bayesian calibration framework to determine the posterior parameter distributions of a hyper-viscoelastic constitutive model using mechanical testing data of brain tissue \cite{teferra2019bayesian}. 

Meanwhile, several studies have been conducted on the failure probability of materials and structures, but not on the failure of rubber-like materials specifically. Orta and Bartlett studied concrete decks to see the effect of different parameters on the failure probability of concrete deck \cite{orta2015reliability}. An investigation was conducted on failure probability analysis of flexural members strengthened with externally bonded fiber-reinforced polymer composites \cite{wang2010reliability}. Khashaba et al. \cite{khashaba2017fatigue} investigated a failure probability analysis on nano-modified scarf adhesive joints in carbon fiber composites. A study was conducted on failure probability analysis of wind turbine blades under multi-axial loading \cite{dimitrov2017spatial}. Another study investigated life-cycle management for corroding pipelines based on the probability of failure \cite{mishra2019reliability}.

\subsection{Current Research}

First, this work presents a parametric stochastic framework for the Carroll model using Bayesian statistics calibration based on maximum a posterior (MAP) estimation and maximum likelihood estimation (MLE). Next, a non-parametric stochastic constitutive model is developed based on Gaussian Process. We train and predict values of stress based on the experimental data set of mechanical tests conducted on different samples of silicon- and polyurethane-based adhesives. Finally, the probability of failure is analyzed based on the limit state function calculated from the stochastic constitutive model. A sensitivity analysis is employed to show the importance of each parameter in the probability of failure.

 {The main contributions of this work are (1) Bayesian evaluation of constitutive (Carroll) model parameters for hyperelastic behavior of rubber-like materials from two distinct experimental data, (2) the development of confidence bounds for stress-strain curves based on conjugate prior, (3) evaluation of failure probability based on First Order Reliability Method (FORM) and Crude Monte Carlo (CMC) simulation concerning a sensitivity analysis on the effect of model parameters on failure probability}.

The paper is outlined as follows. In section \ref{ET}, experimental tests and their results are presented in detail. In section \ref{PCM}, a parametric stochastic constitutive model based on Bayesian model calibration is mentioned for both cases of maximum prior estimation and maximum likelihood estimation. Next, we discusses GP as a non-parametric model in detail for finding hyperparameters of the kernel-based on the L-BFGS method. Moreover, the probability of failure analysis based on FORM and CMC simulation is explained in section \ref{PF} according to the importance analysis of parameters of the constitutive model. Finally, results are shown in section \ref{result} for two material data sets, and a conclusion is provided in section \ref{conc}.

\section{Experimental Tests}
\label{ET}

In this part, the measurement method for hyperelastic deformation, setup, and techniques are described. A uniaxial test is implemented for two materials, silicon (DC) and polyurethane black (PUB). Four specimens are used for each material.

\subsection{Specimen}
\begin{figure}[H]
\centerline{\includegraphics[width=.6\textwidth]{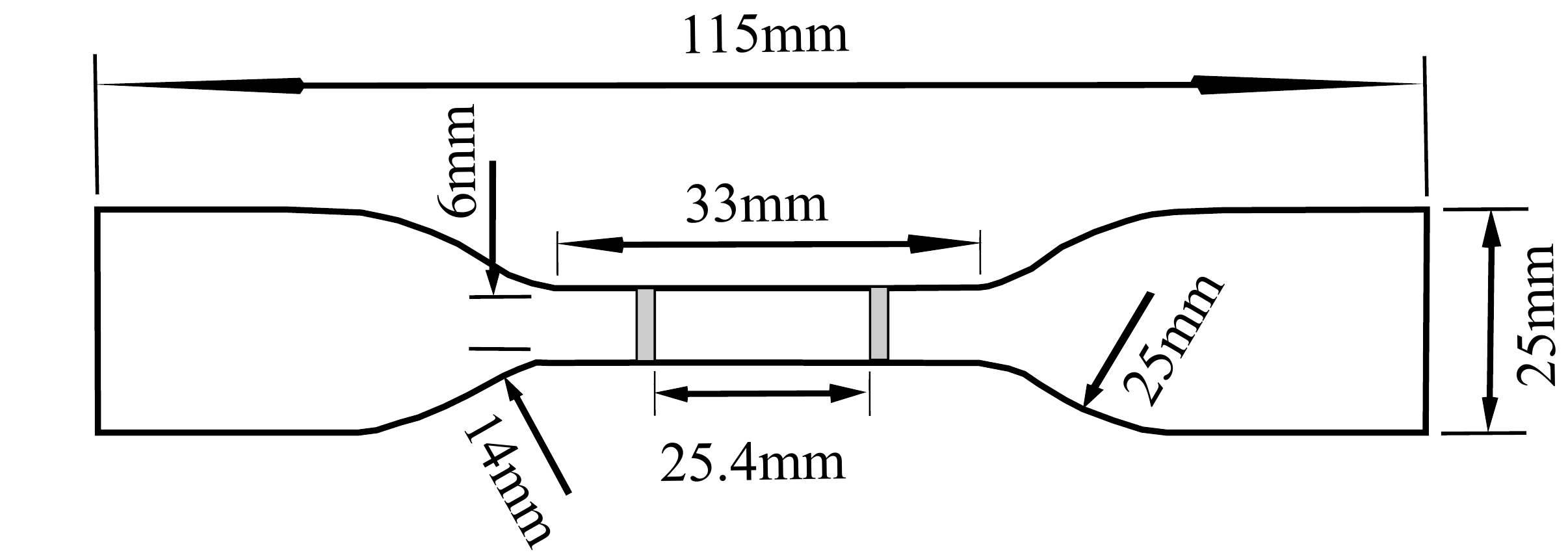}}
\caption{Detailed sample dimensions}
\label{dumbbell}
\end{figure}

From the same batch, each sample had a dumbbell shape. Each sample was cast based on the standard dimension (ASTM D412- Die C) in Fig. \ref{dumbbell}.

\subsection{Mechanical Test}
\begin{figure}[H]
    \centering
   {\includegraphics[width=.95\textwidth]{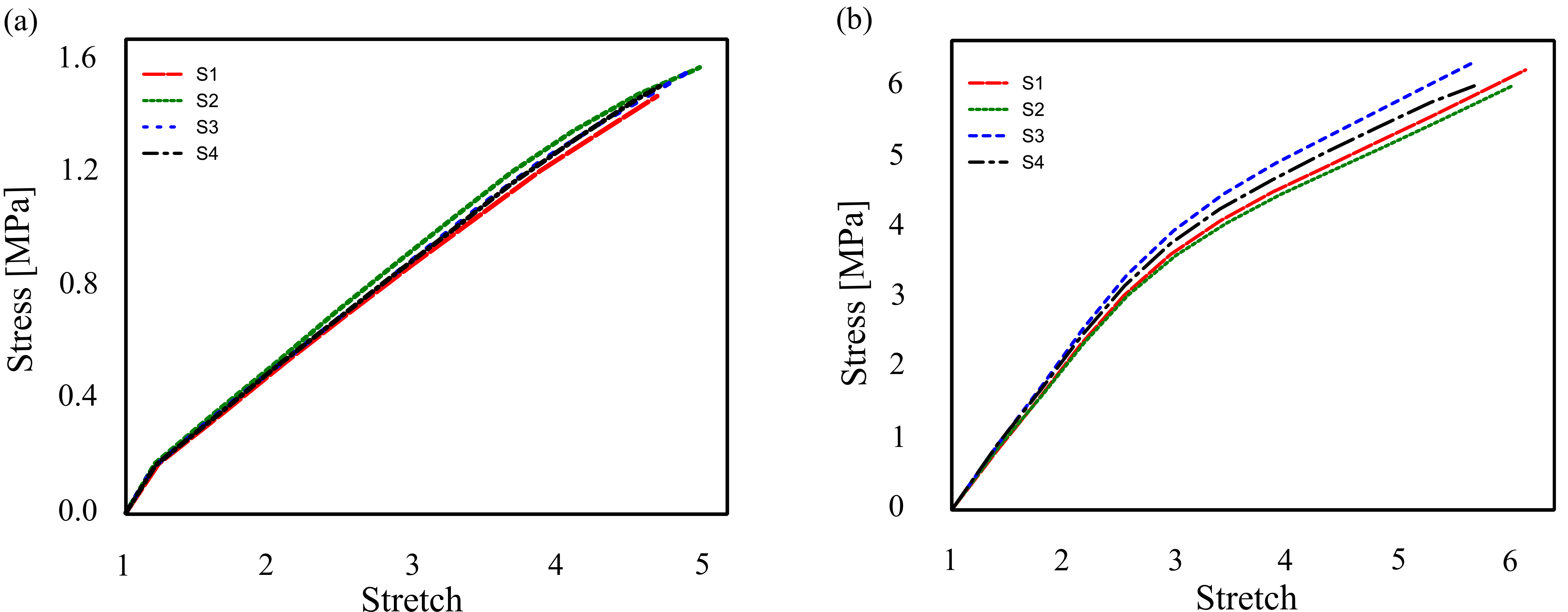}}
    \caption{Stretch-Stress results of mechanical tests for a) DC b) PUB}
    \label{DC_PUB} 
\end{figure}

Quasi-static tensile tests were conducted on a uniaxial universal Testing Machine (TestRecources 311 Series Frame). Samples are clamped between two grips with loading at the rate of $50mm/min$ at room conditions (i.e. $22\pm 2^{\circ}C, 50\pm 3 \% RH$ ). Measurement is conducted by an external extensometer \cite{bahrololoumi2020multi, bahrololoumi2019hydrolytic}. In Fig. \ref{DC_PUB}, stretch-stress curves are depicted for all samples.  As can be seen in this figure, the samples' responses are very close to each other in small deformation. As deformation increases, the evolution of defects in the samples leads to uncertainty in the material's response. 

\section{Parametric Stochastic Constitutive Model}
\label{PCM}

\subsection{Continuum Mechanics}
\label{CM}

Consider that $\tens X$ and $\tens x$ are the reference and current coordinates of an element under deformation  $\tens x= D(\tens X) $ in a body which $D$ is a mapping function and $\tens F= \frac{\partial \tens x}{\partial \tens X}$ known as deformation gradient. We can define right Cauchy-green deformation tensor as $\tens C= \tens F^T \tens F$. $\lambda_{k}$ and $\lambda_{k}^2$, $k=1,2,3$, are eigenvalues of $\tens F$ and $\tens C$, respectively. The principal invariant of $\tens C $ mentioned as

\begin{equation}
    \label{C_f}
    \tens I_1 (\tens C) = tr(\tens C), \quad \quad \quad \quad
    \tens I_2 (\tens C) = \frac{1}{2} ((\tens I_1 (\tens C))^2 - tr(\tens C^2)), \quad \quad \quad \quad
    \tens I_3 (\tens C) =\det (\tens C).
\end{equation}

The strain energy function can be defined as

\begin{equation}
    \label{energy_f}
    \Psi = \Psi(\tens C)=\Psi (\tens I_1,\tens I_2, \tens I_3).
\end{equation}

First Piola-Kirchhoff stress tensor $\tens{P}$ can be written as

\begin{equation}
    \label{stress_f}
    \tens{P}= \frac{\partial{\Psi}}{\partial {\tens{F}}}-{p}{{\tens{F}}^{-T}}.
\end{equation}

For incompressible elastomers, $\det \tens{{F}}=1$ and p is a Lagrange multiplier that arises from the assumption of incompressibility. Here, the hyperelastic response is characterized by Carroll model \cite{carroll2011strain}. The strain energy function and uni-axial stress of the Carroll model is defined by

\begin{equation}
    \label{caroll_f}
    \Psi = W_1\tens I_1 +W_2\tens I_1^4 +W_3 \sqrt{\tens I_2},
\end{equation}

\begin{equation}
    \label{caroll1_f}
    P^{UT} = [2W_1+8W_2[2\lambda^{-1}+\lambda^2]^3+W_3[1+2\lambda^3]^{-\frac{1}{2}}][\lambda-\lambda^{-2}],
\end{equation}
where $\lambda$ is the principal stretch in the uniaxialn tesnile loading, and $W_i \left.\right|_{i=1..3}$  are model parameters.

\subsection{Stochastic Modeling}
\label{SM}
In UQ analysis, uncertainty sources can be categorized because of (1) system inputs uncertainty (e.g. boundary conditions) (2) model form and parameter uncertainty due to lack of knowledge (3) computational uncertainty (e.g. simplification) (4) physical testing uncertainty (e.g. measurement uncertainty). Model uncertainty is the hardest one among these sources due to limited knowledge and inaccurate experimental data. Let us first describe the steps required for the calibration of a constitutive model as a stochastic model. Any arbitrary constitutive model should connect deformation input  $\lambda$ to stress output $p$ through the use of some parameters $\vect W$ as follow
\begin{equation}
    \label{f}
    \vect P = f(\boldsymbol{\lambda}; \vect W)+\epsilon, \quad \textit{with} \quad 
    f(\boldsymbol{\lambda}; \vect W)= \sum_{j=1}^m W_j \phi_j(\lambda), \quad \textit{and} \quad 
     \mathcal{P}( \epsilon)=   {N} (\epsilon|0,\sigma^2),
\end{equation}
where  $\phi_j$ are the basis functions, and $W_i$ are the weight parameters gathered in vector $\vect W$.  
Here,  $ {N} (\epsilon|0,\sigma^2)$ represents the Gaussian distribution of noise  around zero with variance $\sigma^2$ which is assumed to represent $ \mathcal{P}( \epsilon)$.
In a stochastic  calibration problem \cite{mihai2018stochastic},  we optimize function  $f(\boldsymbol{\lambda}; \vect W)$ by fitting   $\vect W$ with a dataset $\vect D$ of $n$ observations of $\lambda$, and $p$ as summarized below
\begin{equation}
  \label{obs}
  \vect D= \left\{\left[\lambda_1,P_1\right] ... \left[\lambda_n,P_n\right] \right\}= \left\{\vect P,\boldsymbol{\lambda}\right\}.
\end{equation}
For the Carroll model, which was described in Eq. \ref{caroll1_f}, can be represented by three weight parameters and three basis functions, where $m=3$. As a representative of other constitutive models, the Carroll model is chosen due to its performance in predicting different states of deformation, which has a rational error. In order to apply a stochastic model calibration, we are using Bayesian methodology.

\subsubsection{Bayesian Methodology}
To calculate the joint probability distribution of model parameters that shows uncertainty associated with experimental data, Bayesian model calibration is created \cite{box2011bayesian}. Compared to the least square method that determines the best parameters for fitting and does not provide any information regarding parameters' probability, the Bayesian approach can show the model's uncertainty with stochastic parameters. This method is based on the Bayes conditional rule of probability, which is written as 

\begin{equation}
    \label{bayes_f}
    \mathcal{P}(\vect W|\boldsymbol{D},M)= \frac{\mathcal{P}(\boldsymbol{D}|\vect W,M) \mathcal{P}(\vect W|M)}{\mathcal{P}(\boldsymbol{D})},
\end{equation}
which $\vect W$ is the vector of unknown model parameters, and $M$ is the chosen model, namely, the Carroll model. $P(\vect W|M)$ is the prior joint distribution and shows the degree of belief to the parameters before we know the data. $\mathcal{P}(\boldsymbol{D}|\vect W,M)$ is the likelihood joint distribution which describes the observation probability of what we have observed, and $\mathcal{P}(\vect W|\boldsymbol{D},M)$ is the posterior distribution. Here, $\mathcal{P}(\boldsymbol{D})$ is a normalizer, given as follow

\begin{equation}
    \label{norm_f}
    \mathcal{P}(\boldsymbol{D}) = \int_{\vect W} \mathcal{P}(\boldsymbol{D}|\vect W,M)\mathcal{P}(\vect W|M)d\vect W.
\end{equation}
For parameter estimation, the marginal likelihood $ \mathcal{P}(\boldsymbol{D})$ does not affect the value of the weight parameters {\vect W}, so it is often considered as a normalization constant. In the absence of any information, the prior probability of the parameters can be assumed to be a Gaussian distribution on the parameter space. It is one way to illustrate our prior ignorance about the weight parameters {\vect W}.

 There are two important ways for model calibration, Maximum Likelihood Estimation (MLE) and Maximum a Posterior (MAP) estimation are explained in the next sections in detail \cite{burr2004bayesian}.
 
{One important aspect of Bayesian methodology is the selection of prior. Different selection process have been introduced for choosing prior such as using right Haar measurement, Jeffreys prior} \cite{robert2007bayesian}, reference priors \cite{berger1992development}, Maxent priors \cite{jaynes2003probability}, conjugate priors \cite{vila2000bayesian}.{ Conjugate priors, unlike other methods, lead to the specific family of distributions for posterior distributions. In this study, we choose prior from the Gaussian family to make the integration of Bayes rule simpler compared to other methods that are problematic and not practical for feeding failure probability analysis. This selection does not affect results significantly because new observation leads to updating of prior distribution after each step. Meanwhile, model selection strategies are manifold in Bayesian methodology. Bayesian methods using Bayes factor, frequentist methods, and Bayesian Information Criterion (BIC) are the most popular approaches. The Bayesian approach has some advantages over frequentist methods. First, it is easier to interpret the posterior probabilities of the model and the Bayes factor as the odds of one model over the other. Second, the Bayesian approach is consistent in the sense that it guarantees the selection of the true model if it is part of the candidate model set under very mild conditions} \cite{berger2001objective, berk1966limiting}. Accordingly, our Priori can be given as
\begin{align}
	\mathcal{P}(\vect W;\alpha)  = \mathcal{N}(\vect W|\vect 0,\frac1\alpha\tens I), 
\end{align}
where the initial assumption is to distribute the weights around zero, $\mathcal{N}(\vect W|\vect 0,\alpha^{-1}\tens I)$, with precision parameter $\alpha$\cite{tipping2001sparse, chen2009bayesian} for MAP that serves as a regulatory index to prevent overfitting and with precision  parameter $\alpha$ equal to $1$ for MLE. The precision parameter is associated with the uncertainty over values of $\vect W$.

\subsubsection{Maximum Likelihood Estimation (MLE)}
The concept of MLE seeks a probability distribution that ‘‘most likely’’ regenerate the observations. In other words, it seeks a weight parameter vector, $\vect W_{MLE}$ which   maximizes the likelihood function   $ \mathcal{P}(\vect W|\boldsymbol{D},M)$.  MLE estimates, $\vect W_{MLE}$, may not exist nor be unique. To reduce computational costs, $\vect W_{MLE}$
 is often obtained by maximizing the log of the likelihood function,  $ \ln \mathcal{P}(\vect W|\boldsymbol{D},M)$, which has a significantly slower growth rate. In essence, since both the likelihood function and its log function are monotonically related to each other, $\vect W_{MLE}$ should maximize both. Assuming  log-likelihood to be differentiable,  MLE principal yields the following conditions
\begin{align}
    \label{MLE_basics}
\left.	\frac{\partial \ln \mathcal{P}(\vect W|\boldsymbol{D},M)}{\partial \vect W}\right|_{\vect W=\vect W_{MLE}}=0, \qquad
\left.	\frac{\partial^2    \ln \mathcal{P}(\vect W|\boldsymbol{D},M)}{\partial \vect W^2}\right|_{\vect W=\vect W_{MLE}} < 0, \qquad
\end{align}
where the second condition ensures the convexity of the function at the optimum $\vect W_{MLE}$.
Rewriting Eq. \ref{bayes_f} with respect to stress and deformation, the posterior summarizes our state of knowledge after observing constitutive data if we know the noise variance $\sigma^{2}$. In view of Eq. \ref{bayes_f}, the posterior is given as

\begin{equation}
    \label{posterior_f}
    \mathcal{P}(\vect W|\boldsymbol{\lambda},\vect P; \sigma^2) = \frac{\mathcal{P}(\vect P|\boldsymbol{\lambda},\vect W; \sigma^2)  \mathcal{P}(\vect W)}{\mathcal{P}(\vect P|\boldsymbol{\lambda})},  \qquad
		 \mathcal{P}(\vect P|\boldsymbol{\lambda})={\int \mathcal{P}(\vect P|\boldsymbol{\lambda},\vect W; \sigma^2)  \mathcal{P}(\vect W) d\vect W},
\end{equation}
where $\mathcal{P}(\vect p|\boldsymbol{\lambda})$ is the marginal likelihood of producing the experimental dataset $\vect D$. Assuming likelihood and priori to be Gaussian, we can write the likelihood function with respect to Eq. \eqref{MLE_basics} as 
\begin{align}
\label{likelihood123}
	%
	\mathcal{P}(\vect P|\boldsymbol{\lambda},\vect W; \sigma^2) = \exp\left(-\frac{1}{2\sigma^2} \rVert{\tens{\Phi}\vect W-\vect P}\rVert^2\right)
	%
\end{align}
 where $\tens{\Phi}\in \mathbb{R}^{n \times m}$ is a matrix with values of basis functions distributed over observation points $\vect D$ such that $\Phi_{i,j}= \phi_i(\lambda_j)$. In MLE approach, one can find the weight parameters, $\vect W_{MLE}$, by derivation from Eq. \ref{likelihood123} 
\begin{equation}
    \label{posterior1_f}
    \nabla \mathcal{P}(\vect P|\boldsymbol{\lambda},\vect W; \sigma^2) = 0 \quad \Rightarrow \quad
    \vect W_{MLE}=\left(\tens{\Phi}^T \tens{\Phi} \right)^{-1} \tens{\Phi}^T \vect P, 
				\qquad   
				\sigma_{MLE}^2=\frac{\rVert{\tens{\Phi} \vect W_{MLE} - \vect P}\rVert^2}{n}.
\end{equation}
The posterior function is consequently derived as a PDF function of multivariate distribution $\mathcal{N}$ as
	\begin{equation}
    \label{pred1}
   \mathcal{P}(\vect P|\boldsymbol{\lambda}, \vect W; \sigma^2) = \mathcal{N} (\vect P|\vect W_{MLE}^T \boldsymbol{\Phi}(\boldsymbol{\lambda});\sigma_{MLE}^2).
\end{equation}
The posterior function allows us to make a probability distribution for a new target constitutive values $\lambda^*$, and $\sigma^*$ based on the optimized weight parameters $\vect W_{MLE}$ using $\mathcal{P}(\vect P^*|\boldsymbol{\lambda^{*}}, \vect W_{MLE}; \sigma)$,where the median $m(\lambda)$, lower bound $l(\lambda)$, and upper bound $u(\lambda)$,  for any new target deformation can be calculated as 
\begin{equation}
    \label{}
    m(\lambda) = \sum_{j=1}^m W_{{MLE}:j} \phi_j(\lambda), \quad \quad \quad l(\lambda) \approx m(\lambda)-2\sigma_{MLE}, \quad \quad \quad u(\lambda) \approx m(\lambda)+2\sigma_{MLE}.
\end{equation}

\subsubsection{Maximum A Posteriori (MAP) Estimation}
In this method, in order to find parameters, the measurement process is modeled using the posterior. We believe that our measurement is around the model prediction, but it is contaminated by Gaussian noise. So, we have the same likelihood here Eq. \ref{likelihood123}. The difference between this method with the last method is, here, we maximize posterior. {This method is very similar to MLE, with the addition of the prior probability over the distribution and parameters. In fact, if we assume that all values of weights are equally likely because we do not have any prior information, then both calculations are equivalent. Thus, both MLE and MAP often converge to the same optimization problem for many machine learning algorithms because of this equivalence. This is not always the case. If the calculation of the MLE and MAP optimization problem differ, the MLE and MAP solution found for an algorithm may also differ.}
We model the uncertainty in model parameters using a prior. In the MAP approach, the likelihood function in Eq. \eqref {posterior_f}
 will be calculated differently based on conjugate prior and given as
\begin{align}
\label{mapder}
	%
	\mathcal{P}(\vect P|\boldsymbol{\lambda},\vect W; \sigma) = \exp\left(-\frac{1}{2\sigma^2} \rVert{\tens{\Phi}\vect W-\vect P}\rVert^2-\frac{\alpha}{2}\rVert{\vect W}\rVert^2\right),
	%
\end{align}
where one can find the optimized weight parameters that are gathered in $\vect W_{MAP}$, by derivation from Eq. \ref{mapder}
\begin{equation}
    \label{posterior2_f}
     \nabla \mathcal{P}(\vect P|\boldsymbol{\lambda},\vect W; \sigma^2) = 0 \quad \Rightarrow \quad
    \vect W_{MAP}=\argmax \log \mathcal{P}\left(\vect P|\boldsymbol{\lambda},\vect W; \sigma\right)=\tens{S}_M \tens{\Phi}^T \vect P, 
				\qquad  \tens{S}_M= \left(\alpha \tens I+  \frac{1}{\sigma^2}\tens{\Phi}^T \tens{\Phi} \right)^{-1}.
\end{equation}
The posterior function  is consequently derived as
\begin{align}
    \label{posterir_2}
    \mathcal{P}(\vect W|\boldsymbol{\lambda},\vect P)&=  \mathcal{N}(  \vect W | \vect W_N, \tens{S}_N  )= \det (2\pi \tens{S}_N)^{-\frac{1}{2}} \exp \left( -\frac{1}{2}(\vect W-\vect W_N)^T \tens{S}_N^{-1} (\vect W - \vect W_N) \right),
\end{align}
where  $\vect{W}_N$ is the mean vector, $\tens{S}_N$ is covariance matrix, and for a Gaussian posterior, $\vect W_{MAP} = \vect W_N$, and $\tens S_M = \tens S_N$. The posterior function  allows us to make a probability distribution for a new target deformation values   $\lambda^{*}$  based on the optimized weight parameters $\vect W_{MAP}$ as
\begin{equation}
    \label{pred}
   \mathcal{P}(\vect P^{*}|\boldsymbol{\lambda^{*}}, \vect W_{N}, \tens S_N) = \mathcal{N} (\vect P^{*}|\vect W_{N}^T \boldsymbol{\phi}(\boldsymbol{\lambda^{*}}),\tens S_N),
\end{equation}
where the median $m(\lambda)$, lower bound $l(\lambda)$, and upper bound $u(\lambda)$,  for any new target deformation can be calculated as 
\begin{equation}
    \label{}
    m(\lambda) = \sum_{j=1}^m W_{{MAP}:j} \phi_j(\lambda), \quad \quad \quad l(\lambda) \approx m(\lambda)-2\sigma, \quad \quad \quad u(\lambda) \approx m(\lambda)+2\sigma.
\end{equation}


\textbf{Non-parametric Stochastic Constitutive Model:}
 In this part, a non-parametric model is proposed that exhibits not only the behavior of elastomers but also uncertainty in the model. The reason for proposing this method is that we have a general study which can be applicable from simple mapping to complicated mapping. Gaussian processes (GP) take a non-parametric approach to model selection. Compared to Bayesian linear regression, GP is more general because the form of the classifier is not limited by a parametric form. GP can also handle the case in which data is available in different forms, as long as we can define an appropriate covariance function for each data type. Here, a GP investigation is conducted to ensure even complex mapping, which standard regression cannot capture, can be calibrated. However, the computational cost is more than a parametric regression. 

\paragraph{Gaussian Process (GP)}

A GP is a probability measure over functions such that the function values at any set of input points have a joint Gaussian distribution \cite{rasmussen2003gaussian, tamhidi2020conditioned}. This property, and the fact that for any set of observations with joint Gaussian distribution, the distribution of a subset conditioned on the rest is also Gaussian, enables making predictions at an unknown point ($\boldsymbol{\lambda^*}$) based on previous observations $\vect D$.
A GP on a model can be written as 

\begin{equation}
    \label{GP_f}
    \mathcal{P}(\vect P) = \mathcal{G_P}(\vect P|\boldsymbol{\mu}; \tens K(\boldsymbol{\lambda}, \boldsymbol{\lambda})),
\end{equation}
where $\boldsymbol{\mu}$ is mean function, $\tens K$ is a symmetric matrix (kernel), which stores the covariance between every pair of inputs in $\boldsymbol{\lambda}$ and depends on a set of hyperparameters $ \boldsymbol{\theta}$. $\mathcal{P}(\vect P)$ denotes our beliefs about $\vect P$. The mean function indicates the central tendency of the function. Assuming no particular knowledge about the trend of the function, we pick a zero mean function. The choice of the covariance function models our prior beliefs about the regularity of the function (there is a one-to-one correspondence between the differentiability of the covariance function and samples from the GP probability measure \cite{adler2010geometry}). We choose the squared exponential covariance function

\begin{equation}
    \label{kernel}
    K_{ij}=k(\lambda_i,\lambda_j)= \nu_0 \exp [-\frac{1}{2} \sum_{n=1} \frac{({\lambda_i}^n-{\lambda_j}^n)^2}{l_n}],
\end{equation}
where  $\{\nu_0, l_1, l_2, ..., l_n\} \in \boldsymbol{\theta}$, and ${\lambda_i}^n$ is n-th element of ${\lambda_i}$ from data set. $ {l_n}$ and $\nu_0$ are hyperparameters known as length-scale and output-scale respectively. Consider that a GP prior $\mathcal{G_P}(\vect P|\mu; \tens K)$ is chosen for the constitutive model $M$, and our experimental data set is $D=[(\lambda_i,P_i)]$, where $P_i=M(\lambda_i)+\epsilon_i$ and $\mathcal{P}(\epsilon|\lambda)=\mathcal{N}(\epsilon,0;\sigma^2)$; we can write GP prior as 

\begin{equation}
    \label{prior_f1}
    \mathcal{P}(\vect P|\boldsymbol{\theta})=\mathcal{N}(\vect P|\boldsymbol{\mu}(\boldsymbol{\lambda}|\boldsymbol{\theta}); \tens K(\boldsymbol{\lambda},\boldsymbol{\lambda}|\boldsymbol{\theta})), 
\end{equation}

To fit the hyperparameters, we look for the $\boldsymbol{\theta}$ that maximizes the log-likelihood \cite{lee2020propagation}. Based on Eq. \ref{posterir_2} for a Gaussian distribution, log-likelihood can be written as 

\begin{equation}
    \label{loglikelihood_F1}
    \log \mathcal{P}(\vect P|\boldsymbol{\lambda},\boldsymbol{\theta})= -\frac{(\vect P-\boldsymbol{\mu})^T \tens V^{-1} (\vect P-\boldsymbol{\mu})}{2}-\frac{\log \det \tens V}{2}-\frac{n\log 2\pi}{2}, \quad \quad \text{with} \quad \quad \tens V=\tens K(\boldsymbol{\lambda},\boldsymbol{\lambda}|\boldsymbol{\theta})+\sigma^2 \tens I
\end{equation}

Marginal likelihood indicates the quality of the fit to our training data. To have proper training, we need to maximize log-likelihood to find the best hyperparameters for fitting. In this study, we used L-BFGS method to maximize log-likelihood.
Our goal is prediction of function $M(\boldsymbol{\lambda}^*)$ at some test locations $\boldsymbol{\lambda}^*$. Now, we can calculate mean and covariance functions at $\boldsymbol{\lambda}$ and evaluate multivariate Gaussian distribution. So, we can write joint distribution between the training function $M(\boldsymbol{\lambda})=\vect P$ and the prediction function values $M(\boldsymbol{\lambda}^*)=\vect P^*$. Using Bayes' rule

\begin{equation}
    \label{posterior}
    \mathcal{P}(\tens P^*|\boldsymbol{\lambda}^*, \boldsymbol{D})=\mathcal{N}(\tens P^*|\boldsymbol{\mu_{M|D}}(\boldsymbol{\lambda}^*); \tens K_{M|D} (\boldsymbol{\lambda}^*,\boldsymbol{\lambda}^*)),
\end{equation}
where
\begin{equation}
    \label{}
    \boldsymbol{\mu_{M|D}}(\boldsymbol{\lambda}^*) = \boldsymbol{\mu}(\boldsymbol{\lambda}^*)+\tens K(\boldsymbol{\lambda}^*,\boldsymbol{\lambda})(\tens K(\boldsymbol{\lambda},\boldsymbol{\lambda})+\sigma^2 \tens I)^{-1}(\tens P-\boldsymbol{\mu}(\boldsymbol{\lambda})),
\end{equation}
 and
\begin{equation}
    \label{}
    \tens K_{M|D}(\boldsymbol{\lambda}^*,\boldsymbol{\lambda}^*)=\tens K(\boldsymbol{\lambda}^*,\boldsymbol{\lambda}^*)- \tens K(\boldsymbol{\lambda}^*,\boldsymbol{\lambda})(\tens K(\boldsymbol{\lambda},\boldsymbol{\lambda})+\sigma^2 \tens I)^{-1}\tens K(\boldsymbol{\lambda},\boldsymbol{\lambda}^*).
\end{equation}
where $\tens K(\boldsymbol{\lambda},\boldsymbol{\lambda}^*)$ is the cross covariance between $\boldsymbol{\lambda}$ and $\boldsymbol{\lambda}^*$.

\section {Probability Failure (PF) Calculation of Hyperelastic Materials}
\label{PF}

\begin{figure}[H]
\centerline{\includegraphics[width=1\textwidth]{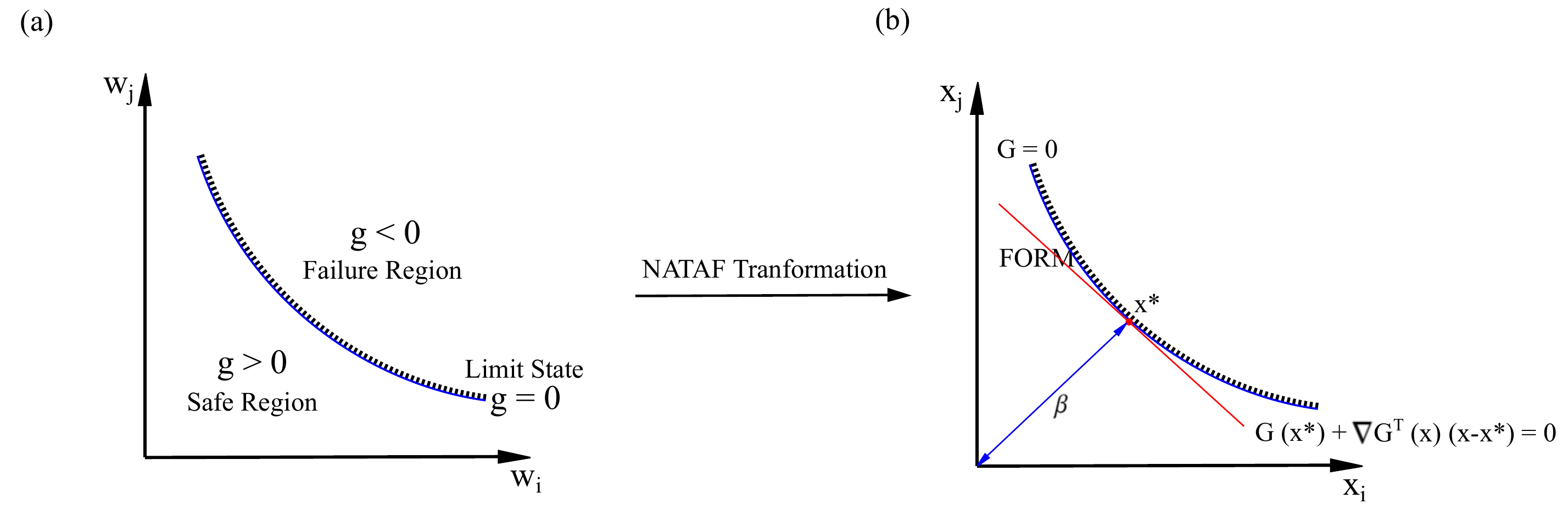}}
\caption{Concept of $\beta$ index with respect to LSF in a) the physical space b) the standard normal space. }
\label{LSF11}
\end{figure}

In failure prediction of hyperelastic materials, uncertainty due to variations in the material matrix, compounding, geometry, and loading conditions can strongly affect the accuracy of predictions. One main challenge in the development of failure prediction engines is the lack of a certain discrete threshold of failure to describe the phenomena through zero and one occurrence. In practice, samples fail over a range of stress-strain amplitudes. This behavior leads to a strong margin of error in the deterministic models of failure. To address this problem, probabilistic models are needed to reproduce the probability of failure (PF). To provide nominal safety indices required for PF, First Order Reliability Method (FORM) and Crude Monte Carlo (CMC) method are popular methods that use analytical and numerical probability integration, respectively. 
Representing the failure stress distribution of the material by the ultimate requirement ($\sigma_U$), one can evaluate PF of this material such that it satisfies $\sigma_U$ using a limit state function (LSF) $g(\vect W)$ as described below. Note that ($\sigma_U$) can be considered as a deterministic value or a probabilistic distribution (see Fig. \ref{LSF11}.a).

\begin{equation}
    \label{LSF}
    g(\vect W) = \sigma_U - P(\vect W), \qquad \Rightarrow \qquad g(\vect W):\left\{\begin{array}{rlr}
> & 0 & \text { Safe region } \\
= & 0 & \text { Limit state } \\
< & 0 & \text { Failure region }
\end{array}\right.
\end{equation}

where $P(\vect W)$ is a constitutive equation of the variables $W_1 , W_2, ..., W_n$.
In failure prediction models, the constitutive model ($P(\vect W)$) can be represented by a probabilistic profile which provides a distribution range of stress for certain strain. The distribution of probabilistic $P(\vect W)$ can be derived  based on the experimental data on stress-strain behaviour of materials.

\begin{figure}[H]
\centerline{\includegraphics[width=1\textwidth]{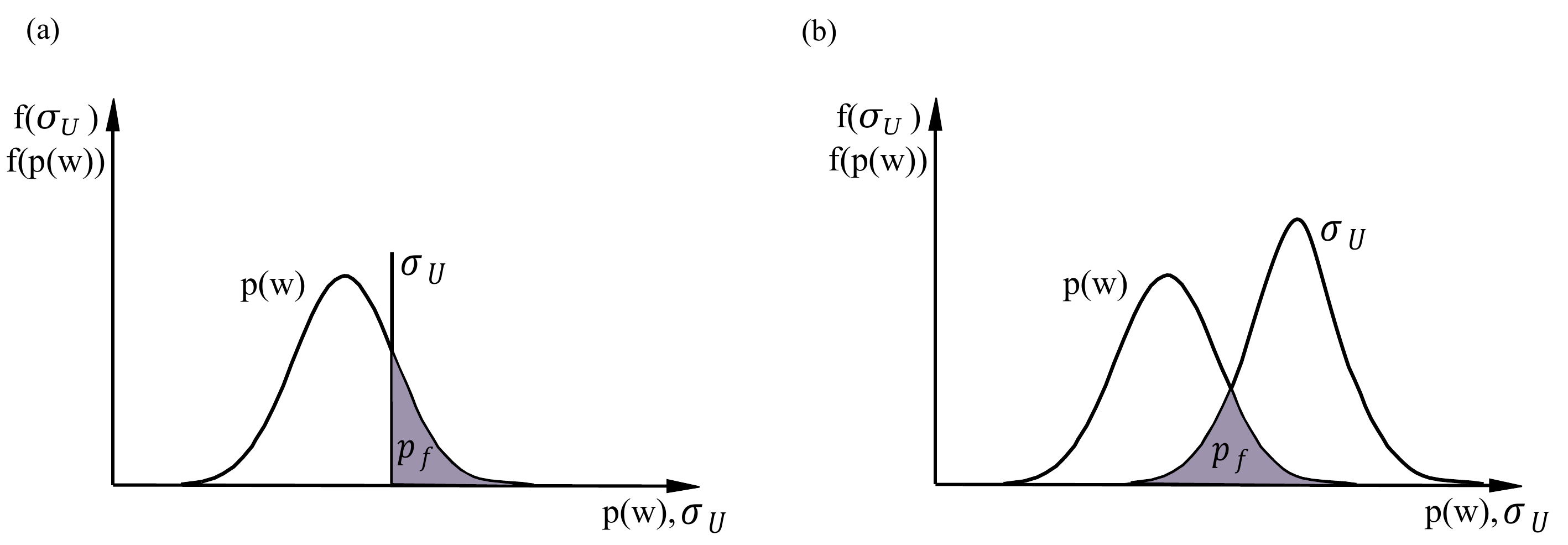}}
\caption{Concept of failure probability with a) a deterministic criterion for ultimate stress b) a probabilistic criterion for ultimate stress. }
\label{LSF1}
\end{figure}

This stochastic constitutive model is estimated from the proposed Bayesian surrogate constitutive model procedure in the past sections (see Eq. \ref{f}).
By using NATAF transformation, random variables which have different distributions can be transformed to the same standard normal space (for computational purposes \cite{lu2014extended}) and then performs a first-order Taylor expansion at the most probable point (MPP), which has the maximum failure probability on the LSF \cite{lebrun2009rosenblatt} (see Fig. \ref{LSF11}.b). To estimate the failure probability (${P_f}$) as the probability that the assembly quality or the limit stated may be violated, the following multi-dimensional integral over the failure domain of the performance function ($g < 0$) should be computed

\begin{equation}
    \label{PF1}
    {P_f} = \mathcal{P} (g < 0) = \int_{g < 0} ... \int f(W_1, W_2, ..., W_n) dW_1 ... dW_n,
\end{equation}
where $f$ is the probability density function, and the number of integrals ($n$) is the number of component dimensions as the random variables. According to Eq. \ref{PF1}, the corresponding integral is calculated in the region that is not desirable ($g < 0$). To solve this problem, the normal standard space is more desirable; therefore, the function $G$ as the LSF in the normal standard space instead of $g$ can be used. To transfer the space of the problem into the normal standard space, the conversion of NATAF transformation \cite{faber2009basics} can be used 

\begin{equation}
    x_i = \Phi^{-1}(f(W_i)), 
\end{equation}
where $x_i$ is the random variable. $f$ and $\Phi^{-1}$ show the corresponding cumulative distribution function (CDF) of the random variable and the inverse cumulative distribution function of the standard normal variable, respectively. Based on the CDF ($\Phi$) concept, failure probability ($P_f$) can be estimated as follows

\begin{equation}
    P_f = \Phi (- \beta),  \quad \quad \beta = \frac{\mu_G}{\sigma_G},  \quad \quad
    \mu_G \cong G(\mu_{x_1}, \mu_{x_2}, ..., \mu_{x_n}), \quad \quad {\sigma_G}^2 \cong \sum_{i=1}^n \sum_{j=1}^n \frac{\partial G}{\partial x_i} \frac{\partial G}{\partial x_j} cov (x_i,x_j),
\end{equation}
where $\beta$ is called the reliability index, representing the shortest distance
from the origin in standardized normal space to the hyperplane or paraboloid, which can be calculated by solving an optimization problem. $\mu_G$ and $\sigma_G$ are the mean and standard deviation of the LSF in the standard normal space. For the FORM  analysis, the limit state function can be described by a linear approximation at the design point ($\vect x^*$), which is the closest point on $G(\vect x) \cong 0$

\begin{equation}
    G(\vect x) = G(x^*) + \nabla G^T(\vect x)(\vect x - \vect x^*), \quad \quad \vect x^* = min [\vect x | G(\vect x) = 0], 
\end{equation}
where $\beta = ||\vect x^* ||$. To find the design point (MPP), the improved Hasofer-Lind-Rackwitz-Fiessler (iHLRF) method can be used \cite{santosh2006optimum}. An arbitrary point $(\vect x_m)$ can be considered as a candidate of the design point in the iteration $m$. The new candidate design point ($\vect x_{m+1}$)) can be obtained as follows 

\begin{equation}
    \vect x_{m+1} = \vect x_m + \delta_m . d_m,  
\end{equation}
where $m$ is the iteration counter, $\delta_m$, and $d_m$ are the step search and the search direction at the $m$ th iteration, respectively. The proper search direction ($d_m$) and the step search ($\delta_m$) for carrying out the search algorithms to find the design point are presented as follows

\begin{equation}
    d_m = (\frac{G(\vect x_m)}{|| G(\vect x_m) ||}+ {{\alpha}^T} \vect x_m) \alpha - \vect x_m,
\end{equation}
where $\alpha = \frac{\nabla G(\vect x_m)}{|| \nabla G(\vect x_m)  ||}$, and $\delta_m = a^K$ can be determined according to Armijo rule which $a$ is a positive constant (usually $a=0.5$ ) and $k$ is an integer that can be iteratively increased from zero \cite{santos2012new}. Note that convergence criteria to stop the search algorithm for finding the design point can be expressed as follows. (1) The design point must be close enough to the surface of the limit state ($G=0$)

\begin{equation}
    \frac{G(\vect x_{m+1})-G(x_m)}{G(x_m)} < e_1,
\end{equation}
$e_1$ shows convergence parameter of the stopping criterion, which is considered 0.001 based on literature in this study. (2) The gradient of the surface of the limit state is passed from the source at the last point, which shows that the current point is the closest point to the origin

\begin{equation}
    || \frac{\vect x_m}{|| \vect x_m ||} - (\alpha_m . \frac{\vect x_m}{|| \vect x_m ||})\alpha_m   || < e_2,
\end{equation}
where $e_2$ is convergence parameter as a criteria to stop iteration. It is 0.001 based on literature. After finding the design point ($\vect x^*$), PF is equal to $P_f = \Phi (-||\vect x^* ||)$.

\begin{figure}[H]
\centerline{\includegraphics[width=.4\textwidth]{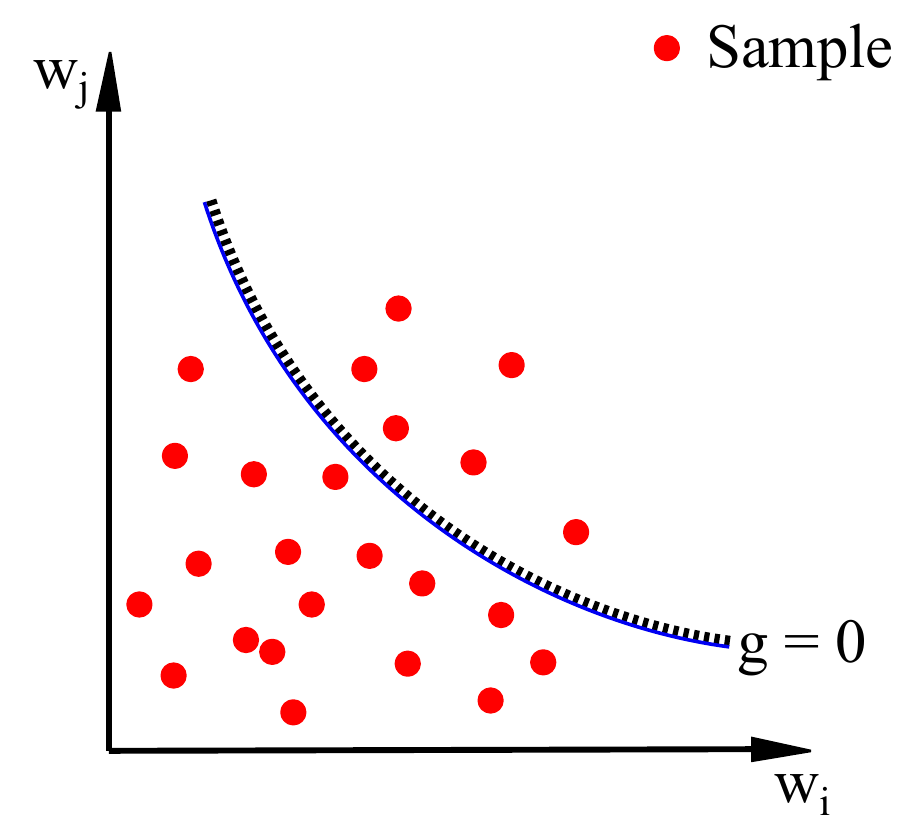}}
\caption{Monte Carlo simulation}
\label{MC}
\end{figure}

\textbf{Crude Monte Carlo (CMC) Simulation:}
In this study, to assess the validity of adopting FORM, CMC simulation is also used concurrently to estimate the limit state probabilities. The results can show the first-order approximation of LSF is an acceptable decision or not. Among the methods of sampling, this method is more commonly used. In this way, what is done is that for random variables $\vect W$, random numbers are generated based on their distribution. Each time a random vector is generated, and then it put in the $g(\vect W)$, when $\vect W < 0$, the sample fails, and vice versa (see Fig. \ref{MC}). Integral of failure probability can be written as 

\begin{equation}
    {P_f} = \mathcal{P} (g < 0) = \int_{g < 0} ... \int f(\vect W) d \vect W=  \int_{-\infty}^{+\infty} ... \int_{-\infty}^{+\infty} I(\vect W)f(\vect W) d \vect W= \frac{1}{N} \sum_{i=1}^N I(W_i),
\end{equation}
where $ N=\frac{1}{\delta_{P_f}^{2}}\left(\frac{1-P_{f}}{P_{f}}\right) $ is the number of simulation, and $\delta_{P_f} = 0.05$ is an common value in the literature \cite{lemaire2013structural}.

For understanding the procedure of this research, the below flowchart shows the steps of this study in summary.

\begin{figure}[H]
\centerline{\includegraphics[width=.7\textwidth]{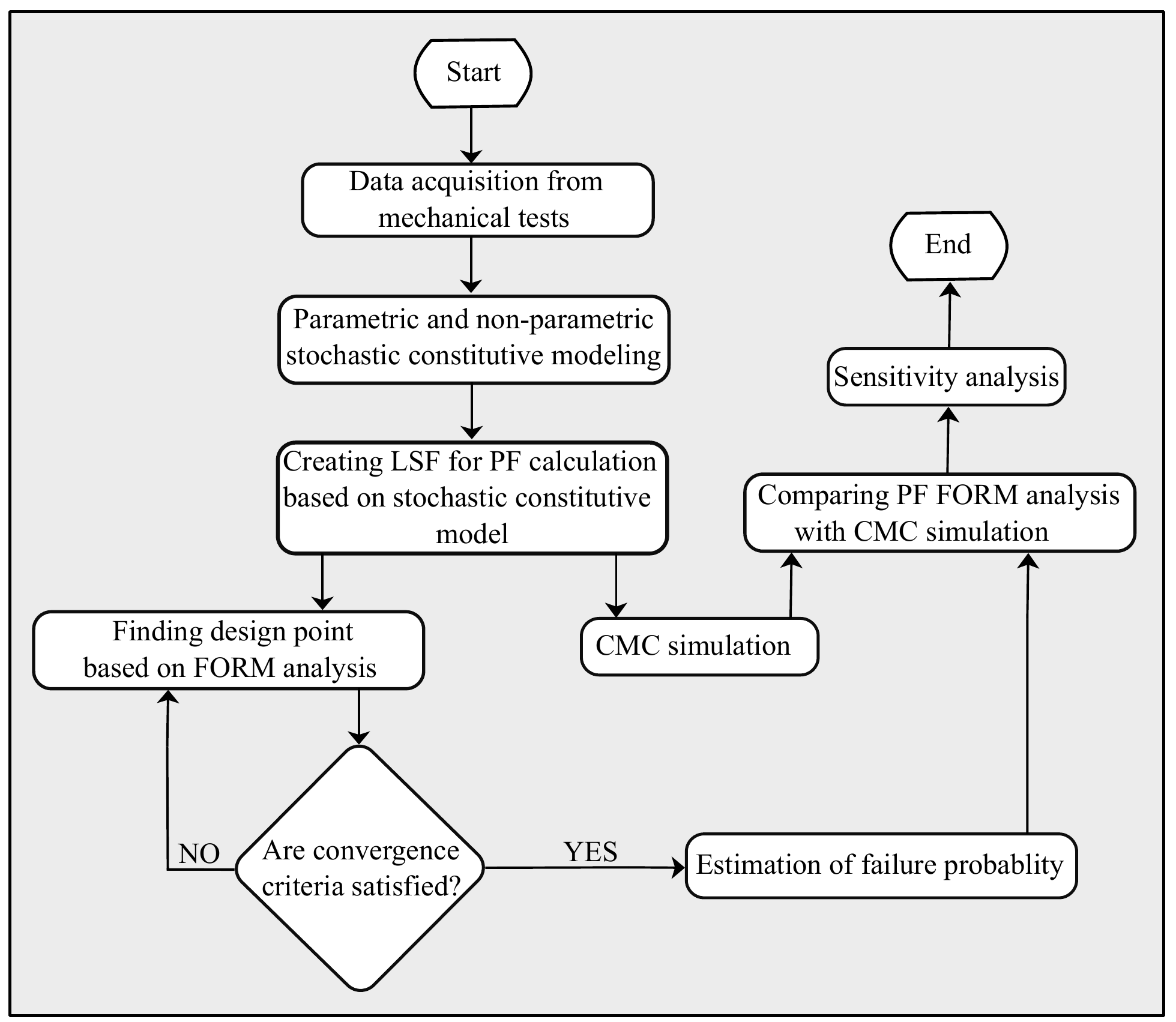}}
\caption{Flowchart of steps of conducted study}
\label{flowchart}
\end{figure}

\paragraph{Sensitivity Analysis}

An importance vector is a computational tool that determines the relative importance of the various parameters involved in the reliability analysis.It has previously been shown that the limit state function is approximated around the design point $G(\vect x) = ||\nabla G(\vect x^*)|| (\beta-\alpha^T \vect x)$. Variance can be computed as follows

\begin{equation}
\label{impvec}
   var(G) = \nabla G^T \Sigma_{xx} \nabla G = (-||\nabla G|| \boldsymbol \alpha)^T (-||\nabla G|| \boldsymbol \alpha) = ||\nabla G||^2 ({\alpha_1}^2+{\alpha_2}^2+...+{\alpha_n}^2),
\end{equation}
which $\Sigma_{xx}$ is covariance matrix. Eq. \ref{impvec} means that the contribution of each random variable in the variance of the limit state function ${\alpha_i}^2$ is related to that random variable. So, the larger the $|\alpha_i|$, the related random variable is more effective. If $\alpha_i >0$, the random variable is called the load variable, and if $\alpha_i <0$ is called the resistance variable. So, whenever the FORM analysis is done, concerning the importance vector, one can determine the significance of random variables, which has the greatest interference in the probability of failure. To consider the correlation between the variables, the importance vector $\boldsymbol {\gamma}$ is defined as follows

\begin{equation}
    \boldsymbol{\gamma} = \frac{\boldsymbol{ \alpha j_{x^*,w^*}} \tens D}{||\boldsymbol{ \alpha j_{x^*,w^*}} \tens D||}, 
\end{equation}

where $\tens j_{x,w} = \tens D \tens L$, and $\tens D$ is derivation matrix in $\boldsymbol \sigma = \tens D \tens R \tens D$ which $\tens R$ is correlation coefficient matrix, and $\tens L = chol(\tens R)$.

\section{Results}
\label{result}

\begin{figure}[H]
    \centering
    \centerline{\includegraphics[width=.85\textwidth]{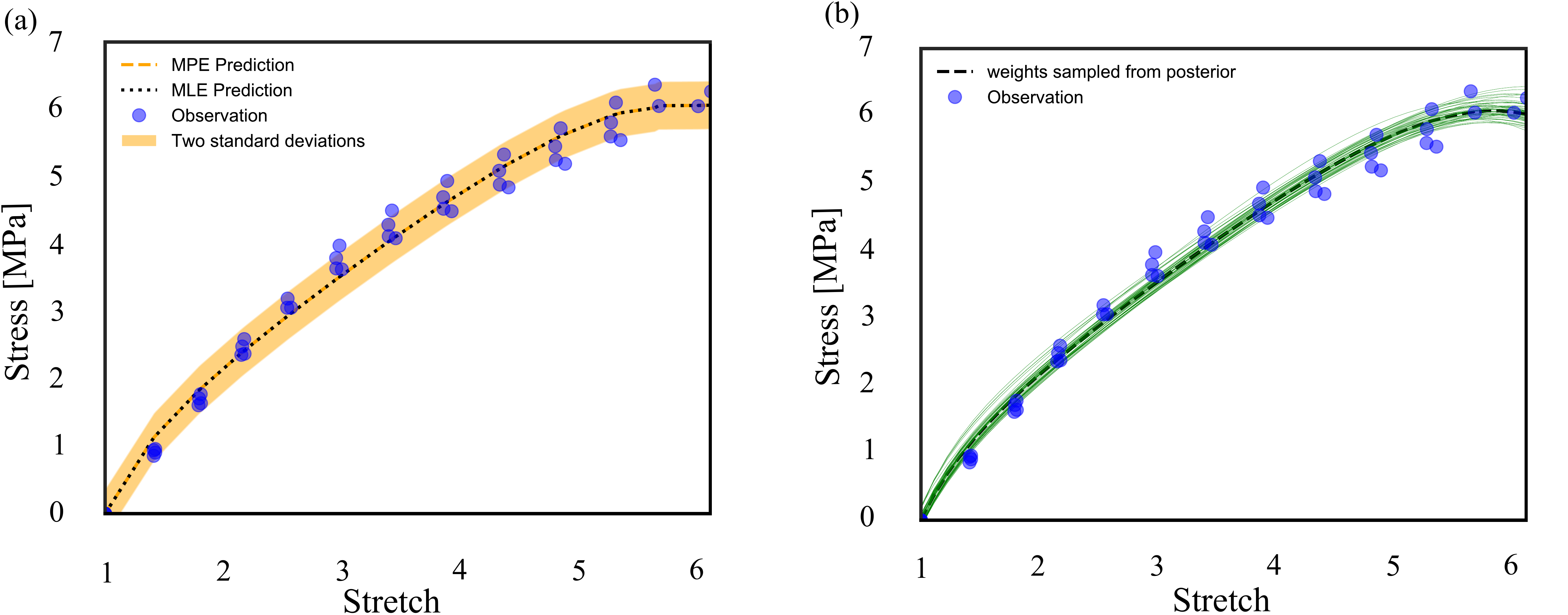}}
    \caption{Model calibration of Carroll model for PUB a) prediction b) plausible models}
    \label{pub_fig}
\end{figure}

\begin{figure}[H]
    \centering
    \centerline{\includegraphics[width=.85\textwidth]{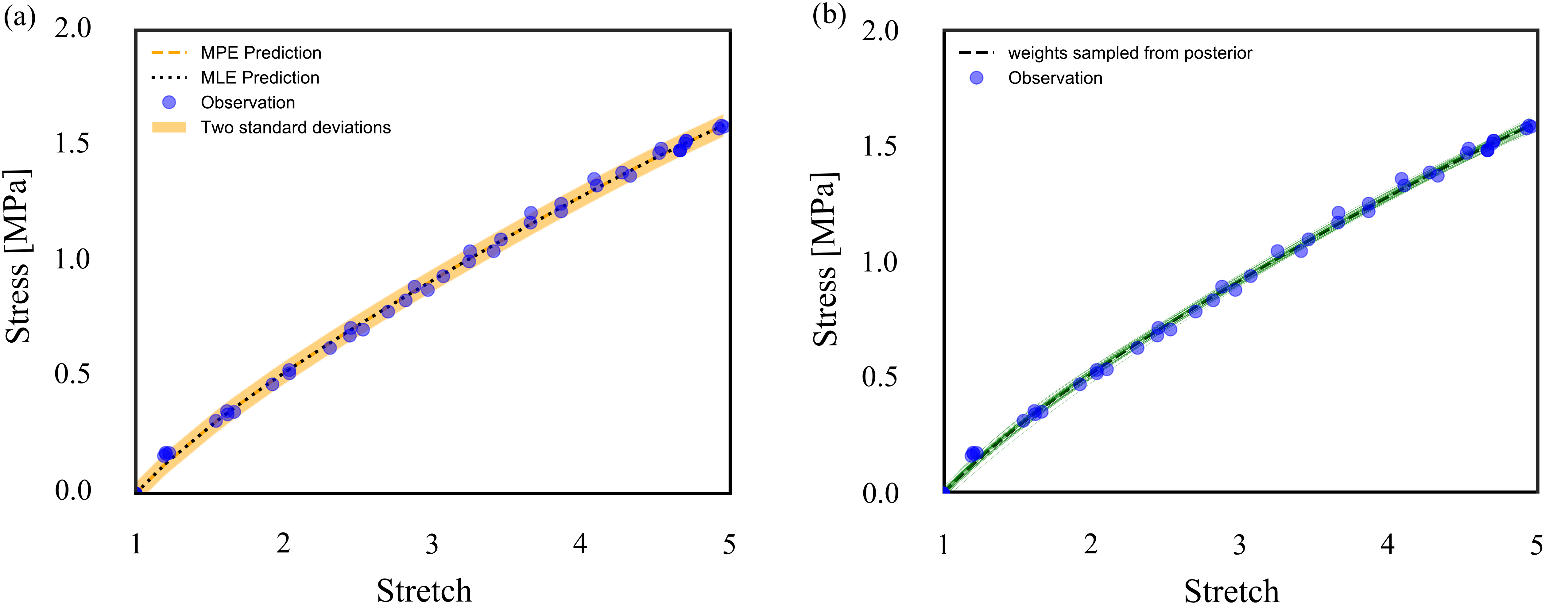}}
    \caption{Model calibration of Carroll model for DC a) prediction b) plausible models}
    \label{dc_fig}
\end{figure}

\paragraph{Parametric Stochastic Constitutive Model}
To show the Carroll model's uncertainty quantification with two observed data sets, model calibration is conducted based on Bayesian regression. MLE and MAP are obtained for the Carroll model for two sets of experimental data. Several plausible models based on MAP are plotted as well. Fig. \ref{pub_fig} and Fig. \ref{dc_fig} show the results for PUB and DC, respectively. Table \ref{PUB.T1} and \ref{DC.T1} show the stochastic parameters of Carroll model for DC and PUB respectively.

\begin{table}[H]
\centering
\caption{Statistical characteristics of Carroll model for PUB}
\begin{tabular}{c c c c c}
\hline 
\multicolumn{1}{l}{Parameters} & \begin{tabular}[c]{@{}c@{}}Statistical\\  distribution\end{tabular} & Mean value & \begin{tabular}[c]{@{}c@{}}Standard\\  deviation\end{tabular} & \begin{tabular}[c]{@{}c@{}}Coefficient \\ of\\ Variation\end{tabular} \\ \hline 
W1                              & Normal                                                              & 0.61025          & 0.01555                                                             & 0.0255                                                                     \\ 
W2                               & Normal                                                              & -5.4944e-7          & 0.9059e-7                                                             & 0.1648                                                                     \\ 
W3                               & Normal                                                              & 0.09649          & 0.23623                                                             & 2.4481                                                                     \\ \hline
\end{tabular}
\label{PUB.T1}
\end{table}

\begin{table}[H]
\centering
\caption{Statistical characteristics of Carroll model for DC}
\begin{tabular}{c c c c c}
\hline 
\multicolumn{1}{l}{Parameters} & \begin{tabular}[c]{@{}c@{}}Statistical\\  distribution\end{tabular} & Mean value & \begin{tabular}[c]{@{}c@{}}Standard\\  deviation\end{tabular} & \begin{tabular}[c]{@{}c@{}}Coefficient \\ of\\ Variation\end{tabular} \\ \hline 
W1                              & Normal                                                              & 0.173568          & 0.002315                                                             & 0.013341                                                                     \\ 
W2                               & Normal                                                              & -8.43e-8          & 3.492e-8                                                             & 0.414199                                                                     \\ 
W3                               & Normal                                                              & -0.206981          & 0.028986                                                             & 0.140046                                                                     \\ \hline
\end{tabular}
\label{DC.T1}
\end{table}

\begin{figure}[H]
    \centering
    \centerline{\includegraphics[width=.85\textwidth]{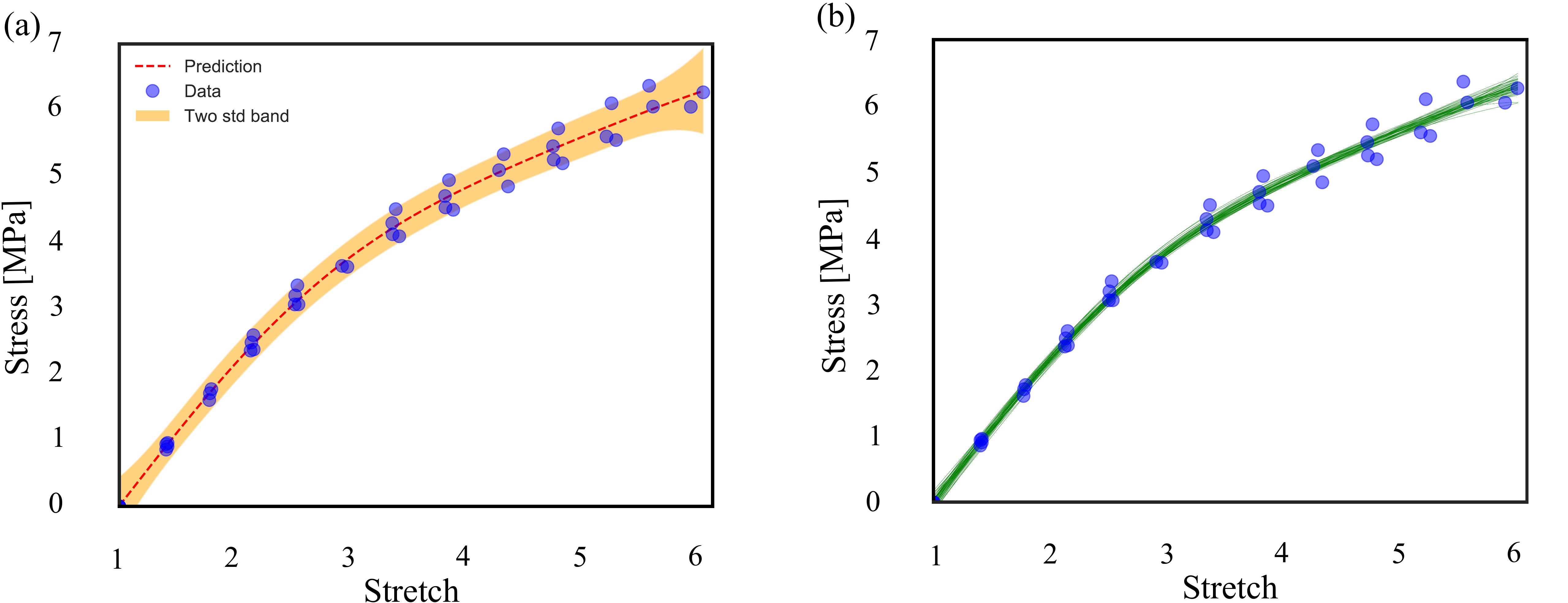}}
    \caption{GP model for PUB a) prediction b) plausible models}
    \label{pub_fig1}
\end{figure}

\begin{figure}[H]
    \centering
    \centerline{\includegraphics[width=.85\textwidth]{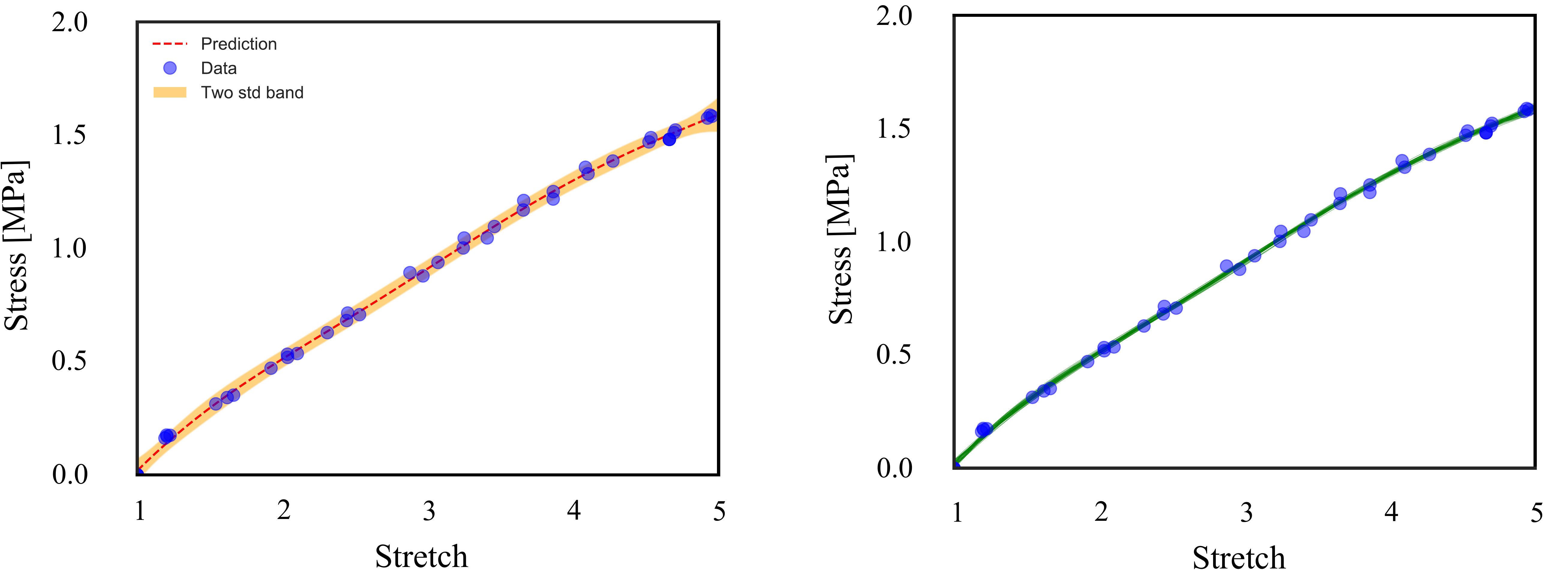}}
    \caption{GP model for DC a) prediction b) plausible models}
    \label{DC_fig1}
\end{figure}

\paragraph{Non-Parametric Stochastic Constitutive Model}
To see the non-parametric model's performance, a GP analysis is conducted on these data sets to find hyperparameters of the kernel in GP. We maximized log-likelihood based on L-BFGS method for two experimental data sets.
Fig. \ref{pub_fig1} and Fig. \ref{DC_fig1} show results for PUB and DC, respectively. Besides, several plausible models are plotted based on obtained hyperparameters. Data is more scatter in larger deformation due to the breakage of some samples and cumulative errors with stretch increasing.

\begin{figure}[H]
    \centering
    \centerline{\includegraphics[width=1.1\textwidth]{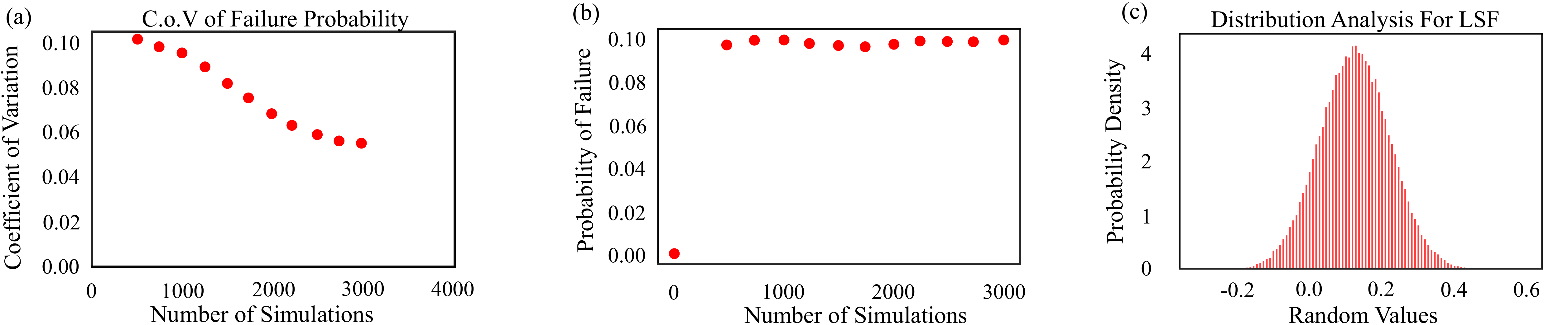}}
    \caption{a) coefficient of variation for failure probability respect to number of simulation b) probability of failure respect to number of simulation c) LSF distribution analysis}
    \label{PUB_F}
\end{figure}

\begin{figure}[H]
    \centering
    \centerline{\includegraphics[width=1.1\textwidth]{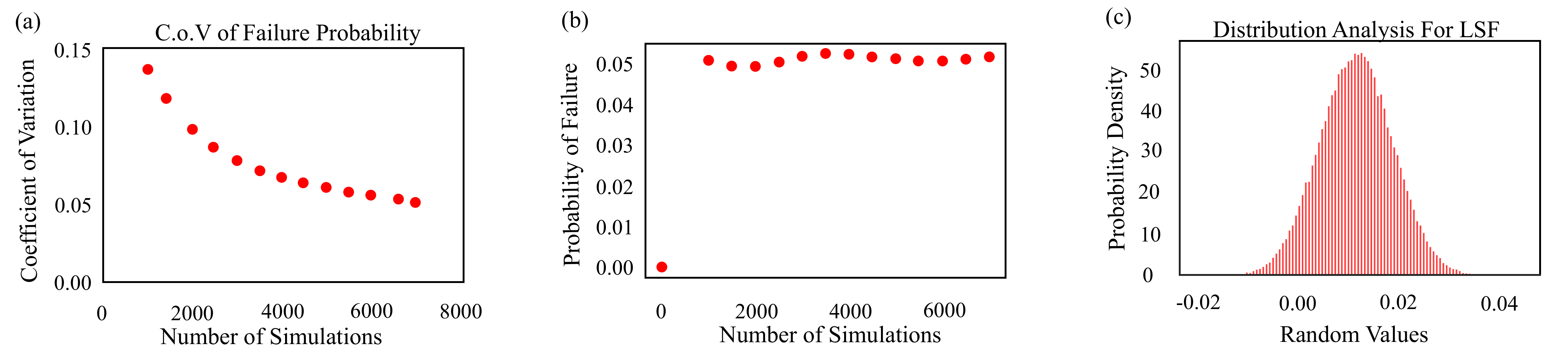}}
    \caption{a) coefficient of variation for failure probability respect to number of simulation b) probability of failure respect to number of simulation c) LSF distribution analysis}
    \label{DC_F}
\end{figure}

\paragraph{Probability of Failure}
For failure analysis, the first step is creating LSF. Based on obtained stochastic constitutive model in the last steps, the LSF can be written as
\begin{equation}
    g(\vect W) = \sigma_U - [2W_1+8W_2[2\lambda^{-1}+\lambda^2]^3+W_3[1+2\lambda^3]^{-\frac{1}{2}}][\lambda-\lambda^{-2}],
\end{equation}
which $\lambda =  5.861 $ for case of PUB and $\lambda = 4.815$ for case of DC. Here, the stretch value is chosen with respect to the mean of stretch data set; however, the selection of stretch for failure calculation depends on how much we stretch the material. $\sigma_U =\mathcal{N}(6.19,0.16) $ and $\sigma_U = \mathcal{N}(5.9,0.237)$ show the value of ultimate strength which is failure criteria for PUB and DC respectively. They have been computed based on the data of four samples at failure points (i.e., a distribution analysis on failure points of four samples in each case). Table \ref{formcmc} shows the results of FORM analysis and CMC for PUB and DC. Also, Fig. \ref{PUB_F} and Fig. \ref{DC_F} Show the details of CMC simulation and distribution of LSF analysis for PUB and DC, respectively.

\begin{table}[H]
\caption{Probability of failure for PUB and DC based on FORM and CMC}
    \centering
    \begin{tabular}{lrr||rrr}
  \hline
  &\multicolumn{2}{c}{PUB}&&\multicolumn{1}{c}{DC}\\
  \cline{2-3}\cline{4-6}
  Method&FORM&CMC&FORM&&CMC\\
  \hline
  $P_f(\%)$ & 10.185 & 9.725 & 5.577 && 5.671\\
  $\beta$ & 1.2710 & 1.2973 & 1.5912 && 1.5829\\
   \hline
   \end{tabular}
    \label{formcmc}
\end{table}

\begin{figure}[H]
    \centering
    \centerline{\includegraphics[width=.8\textwidth]{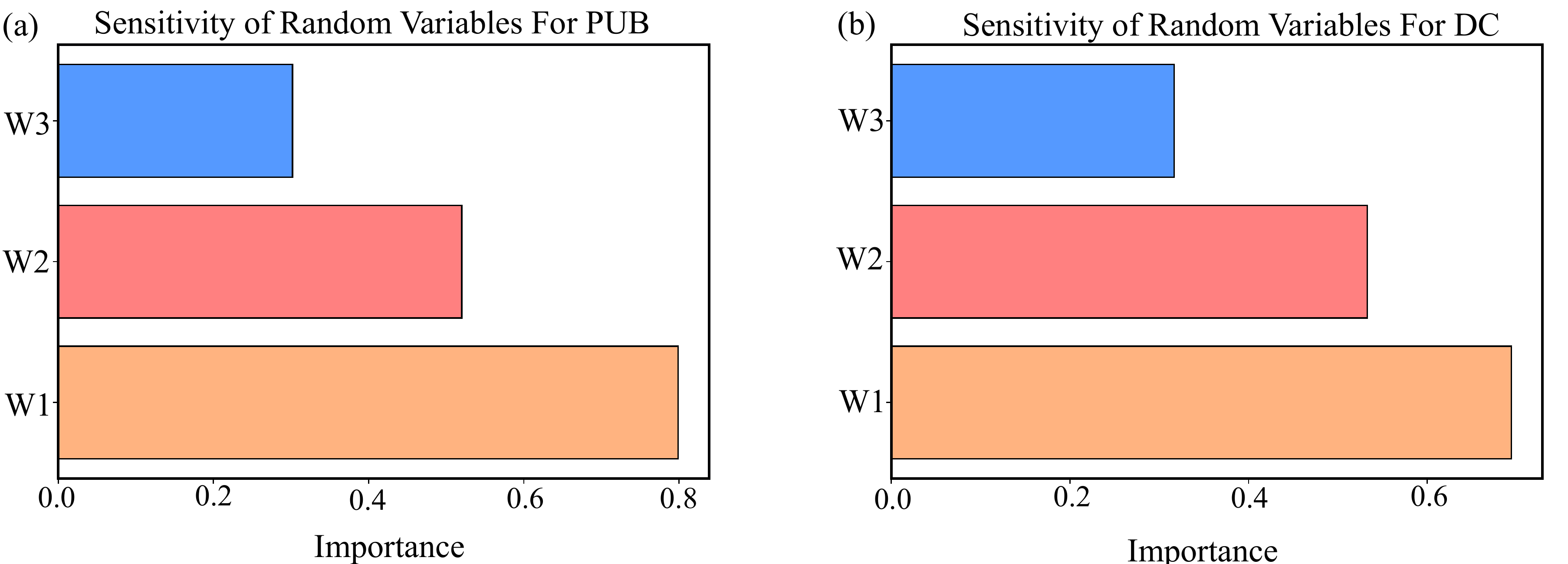}}
    \caption{Importance analysis a) for PUB b) for DC}
    \label{sensitivity}
\end{figure}

Hence, failure probability exhibits if we stretch the material until a specific point, it may fail proportional to a specific probability at that point.
To show the importance of each random variable in failure probability, a sensitivity analysis is conducted. Fig. \ref{sensitivity} shows the results for PUB and DC.

\section{Conclusion}
\label{conc}

In this work, we employed the Bayesian surrogate approach to model a parametric constitutive model (e.g., Carroll model) for the hyperelastic response of DC and PUB based on MLE and MAP estimation. The different mechanical test data sets are considered, which describe uni-axial deformation of DC and PUB. Model selection and parameter estimation should not be considered as separate steps. However, they should be considered as a single framework used to find a reliable model that can predict the quantities of interest. In the next step, the same framework was chosen based on GP as a non-parametric stochastic model. Hyperparameters of the squared exponential kernel were found based on L-BFGS method. This complicated method has the same result in our case study. In the last step, due to uncertainty in the failure point, we conducted a failure probability analysis based on FORM analysis and CMC simulation. The stochastic constitutive model was employed to create LSF for the status of failure. A sensitivity analysis was also investigated to demonstrate the importance of each variable parameter in the failure probability of rubber-like materials. This method is general and can be used for any combination of data and constitutive models.

\appendix
\section{Multivariate Normal Distribution}
\label{app}

The probability density function (pdf) of a multivariate normal distribution, with a random vector ${\boldsymbol{\lambda}}=({\lambda_1},{\lambda_2}, ... ,{\lambda_n})$, mean ${\boldsymbol{\mu}} = ({\mu_1},{\mu_2},...,{\mu_n})$ and positive-definite covariance matrix $\tens S =[{\sigma_{ij}}]$, can be written as 

\begin{equation}
    \label{multi}
    P(\boldsymbol{\lambda}|\boldsymbol{\mu}, \tens S)= \frac{1}{(2\pi)^{\frac{n}{2}} \mid \tens S \mid^{\frac{1}{2}}} \exp [-\frac{1}{2}(\boldsymbol{\lambda}-\boldsymbol{\mu})^T \tens S^{-1} (\boldsymbol{\lambda}-\boldsymbol{\mu})].
\end{equation}

\begin{figure}[h]
\centerline{\includegraphics[width=.7\textwidth]{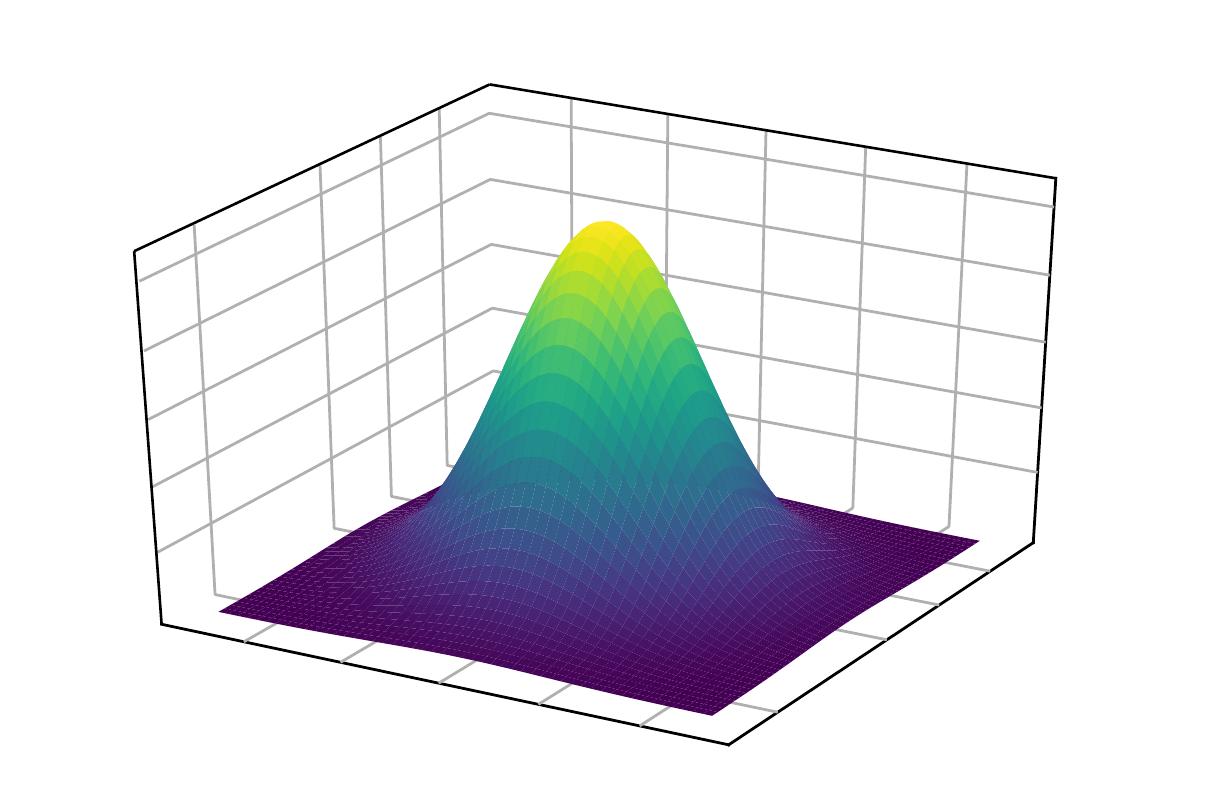}}
\caption{Schematic illustration for bivariate normal distribution}
\label{BND}
\end{figure}

Multivariate normal distribution plays a crucial role in multivariate statistical analysis and has remarkable properties. For example, the summation of some multivariate normal distribution has a normal distribution, and their product has log-normal distribution. If $\boldsymbol{\mu}=0$ and $\tens S=\tens I$, it is called the standard normal distribution. To gain a better insight into the distribution, a bivariate normal distribution is shown in Fig. \ref{BND}.

\pagebreak

\bibliographystyle{main}
\bibliography{main.bib}

\end{document}